\documentclass[12pt]{elsarticle}
\usepackage[english]{babel}

\usepackage{amsmath}
\usepackage{amssymb} 
\usepackage{stmaryrd}
\usepackage{mathtools}

\usepackage{tikz}
\usepackage{tkz-graph}
\usetikzlibrary{matrix,arrows}

\newif\ifcomment
\commenttrue
\definecolor{gris}{gray}{0.3}

\newtheorem{theorem}{Theorem}
\newtheorem{lemma}{Lemma}
\newtheorem{proposition}{Proposition}
\newtheorem{corollary}{Corollary}
\newdefinition{definition}{Definition}
\newproof{proof}{Proof}
\newcommand{\tuple}[1]{\langle #1 \rangle}
\newcommand{\Nset}[0]{\mathbb{N}}
\newcommand{\Bset}[0]{\mathbb{B}}

\newcommand{\partset}[1]{\ensuremath{\textbf{2}^{#1}}} 
\newcommand{\eqdef}[0]{\stackrel{\scriptscriptstyle{\mathrm{def}}}{=}}
\newcommand{\abbrev}[1]{#1, \relax}
\newcommand{\ie}[0]{\abbrev{\textit{i.e.}}}
\newcommand{\eg}[0]{\abbrev{\textit{e.g.}}}
\newcommand{\cf}[0]{\abbrev{\textit{cf.}}}
\newcommand{\wrt}[0]{w.r.t. }

\newcommand{\evo}[0]{\longrightarrow}
\newcommand{\xevo}[1]{\stackrel{#1}{\evo}}
\newcommand{\dom}[0]{\operatorname{\textbf{dom}}\,}
\newcommand{\spectrum}[0]{{\operatorname{\textbf{spx}}\,}}
\newcommand{\mode}[0]{\operatorname{\textbf{md}} \,}
\newcommand{\gsym}[1]{\operatorname{S}_{#1}}
\newcommand{\aut}[0]{\operatorname{Aut}}
\renewcommand{\land}[0]{.}
\renewcommand{\lor}[0]{+}
\renewcommand{\lnot}[1]{\overline{#1}}
\newcommand{\lxor}[0]{\oplus}
 
\begin{document}
\title{Analogous Dynamics of Boolean Network}

\author{Franck Delaplace}
\ead{franck.delaplace@ibisc.univ-evry.fr}
\address{IBISC laboratory, Evry Val d'Essonne University \\ 
 23 boulevard de France 91037 Evry, France}
\cortext[cor1]{Corresponding author}
\begin{abstract}
Different Boolean networks may reveal similar dynamics although their definition differs, then preventing their distinction from the observations. This raises the question about the sufficiency of a particular Boolean network for properly reproducing a modeled phenomenon to make realistic predictions. The question actually depends on the invariant properties of behaviorally similar Boolean networks. In this article, we address this issue by considering that the similarity is formalized by isomorphism on graphs modeling their dynamics. The similarity also depends on the  parameter governing the
updating policy, called the mode. We define a general characterization of the group of isomorphism preserving the mode. From this characterization, we deduce invariant structural properties of the interaction graph and conditions to maintain an equivalence through mode variation.
\end{abstract}

\maketitle{}
\section{Introduction}
Boolean network is a discrete model of dynamical systems based on the evolution of Boolean states. Two components are used for the description: The \emph{evolution function} and the \emph{mode}. The evolution function reckons the state transition with respect to a mode governing the state updating policy. From these components, the dynamics is represented by a labeled transition system, called a \emph{model}, where all the paths/trajectories are computed by iterated applications of the evolution function (Section~\ref{sec:discrete-dynamics}). The mode is a parameter composed of  agent parts
determining the set of agents that could make evolve their state jointly (Section~\ref{sec:discrete-dynamics}).

Boolean network is used in biological modeling for investigating the properties of gene regulation and signal transduction networks \cite{Thomas1991,Garg2007, Demongeot2010, Delaplace2010}. The correspondence between models and observations is based on the assumption that the equilibria of the Boolean dynamics characterize the molecular signatures of observed phenotypes.  However, different models may fit to biological observations. Hence, from the observation standpoint they behave analogously. However,  they  could lead to different predictions for unobserved behaviors due to the lack of facts discriminating them. Consequently, it appears important to delineate the edges of the analogy for the reliability of model prediction.  

The analogy is defined here as an equivalence on trajectories that can be formalized by isomorphism on model: two networks are dynamically equivalent if and only if their model is isomorphic. Finding properties shared by these networks enforces the reliability of the prediction  because the predictions based on these properties remains identical for any network of the class. In this article we characterize the group of isomorphisms preserving the updating policy (\ie the mode) from which we derive the invariant properties of analogous Boolean networks. 

After introducing the formalism of Boolean networks (Section~\ref{sec:discrete-dynamics}), analogy will be formalized by an isomorphism on models (Section~\ref{sec:isomorphic-models}). The contribution is the formal characterization of the family of isomorphisms preserving the mode from which we deduce properties related to network equivalence (Section~\ref{sec:evo-fun}). These properties concern the structural invariance and condition to maintain the equivalence of Boolean networks through mode variations.

\section{Boolean networks}
\label{sec:discrete-dynamics}
Boolean networks defines the discrete evolution of Boolean \emph{state} for a population of agents. More precisely, the state of a set of agents $A$ is a Boolean vector $s \in \Bset^{|A|}, \Bset=\{0,1\}$ indexed by $A$. In the sequel, $s[a]$ denotes the state of agent $a$, and $s[A']$ is a sub-vector of $s$ collecting the states of agents belonging to $A', A' \subseteq A$. For example, given a set of agents $A=\{a_4,a_3,a_2,a_1\}$ and a state $s=(1,0,1,0)$, we deduce that: $s[a_4]=1,s[a_3]=0,s[a_2]=1,s[a_1]=0$ and $s[\{a_4,a_1\}]=(1,0)$. Throughout the article the cardinal of $A$ is $n$ (\ie $n=|A|$). The \emph{ evolution function} is a Boolean function on states, $f:\Bset^{n} \to \Bset^{n}$ used to determine the stepwise evolution of states. $f$ is  defined by a sequence of propositional formulas with the agent names as variables\footnote{We write $a_i$ instead of $s[a_i]$ in the definition of the evolution function.}, for example:
\begin{equation}
\label{eq:ex1}
A = \{a_4,a_3,a_2,a_1\}, \; f=(a_4,a_4 \lor a_2, \lnot a_3, a_2)
\end{equation}
We use the following notations for Boolean operations: $\lor$ is the logical \textsc{or}, $\land$ stands for the logical \textsc{and}, $\lxor$ is the logical \textsc{exclusive or}, and $\lnot{a}$ represents the negation of a formula.

Following the notation used for states, $f_{A'}, A'\subseteq A$ defines a sub-function whose co-domain is restricted to states of $A' $ (\ie $f_{A'}:\Bset^{|A|}\to \Bset^{|A'|}$) determining the state evolution of these agents. $\tilde f_{A'}: \Bset^n \to \Bset^n$ extends $f_{A'}$ by completing the outcome state with the unchanged states of the agents that does not belong to $A'$:
\begin{equation*}
\tilde f_{A'}(s_1) = s_2 \iff s_2[A']=f_{A'}(s_1) \wedge s_2[A\setminus A'] = s_1[A \setminus A'].
\end{equation*} 

An abstraction of the evolution function $f_A$ is given by the \emph{signed interaction graph}, $G_{f_A}$ (Definition~\ref{def:interaction}) that graphically represents the interaction of the agents on the others. The \emph{signs}\footnote{$\{-1,0,1\}$ are respectively represented by $-$, $\pm$, $+$ in the figures.} $\{-1,0,1\}$ indicate a correlation on state variations of the connected agents: either increasing ($+1$) or decreasing ($-1$). $0$ indicates the absence of a monotone relation.

\begin{figure}[ht]
\begin{center}
 \begin{tikzpicture}[scale=1, node distance=2cm]
 \SetVertexNormal[Shape = circle, LineWidth=0.5pt]
\tikzset{LabelStyle/.style = {sloped}}
 \Vertices[Math, x=0, y=0, dir=\EA]{a_1,a_2,a_3,a_4};
 \Edge[ label= + ,labelstyle=above, style={post}](a_2)(a_1)
 \Edge[ label= + ,labelstyle=above, style={post,bend left}](a_2)(a_3)
 \Edge[ label= - ,labelstyle=below, style={post,bend left}](a_3)(a_2)
 \Edge[ label= + ,labelstyle=above, style={post}](a_4)(a_3)
 \Loop[ label= + ,labelstyle=above, style={post},dir=EA, dist=4em](a_4)
\end{tikzpicture}
\end{center}
\caption{Interaction graph of evolution function (\ref{eq:ex1}). }
\label{fig:inet-ex1}
\end{figure}
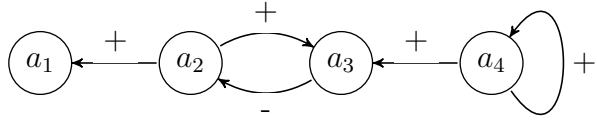

\begin{definition}
\label{def:interaction}
Let $f_A$ be an evolution function, the signed interaction is a relation denoted $\stackrel{\star}{\longrightarrow}$included in $A\times\{-1,0,1 \}\times A$. We define an \emph{interaction} on agents as: 
$$ a_i \longrightarrow a_j \eqdef 
	\begin{array}[t]{l}
	\exists s_1,s_2 \in \Bset^n: s_1[a_i] \neq s_2[a_i] \wedge s_1[A \setminus a_i] = s_2 [A \setminus a_i] \wedge \\
 \quad f_{a_j}(s_1) \neq f_{a_j}(s_2).
	\end{array}
$$
The signed interactions are defined as follows: 
\begin{itemize}
\item $ a_i \stackrel{+1}{\longrightarrow} a_j \eqdef 	\begin{array}[t]{l}
 a_i \longrightarrow a_j \wedge \\
\quad \forall s_1,s_2 \in \Bset^n: s_1[a_i] \leq s_2[a_i] \wedge s_1[A \setminus a_i] = s_2 [A \setminus a_i] \\
\qquad \implies f_{a_j}(s_1) \leq f_{a_j}(s_2).
\end{array}
$ 
\item $ a_i \stackrel{-1}{\longrightarrow} a_j \eqdef
	\begin{array}[t]{l}
 a_i \longrightarrow a_j \wedge \\
\quad \forall s_1,s_2 \in \Bset^n: s_1[a_i] \leq s_2[a_i] \wedge s_1[A \setminus a_i] = s_2 [A \setminus a_i] \\
\qquad \implies f_{a_j}(s_1) \geq f_{a_j}(s_2).
\end{array}$ 
\item $ a_i \stackrel{0}{\longrightarrow} a_j \eqdef 
	\begin{array}[t]{l}
 a_i \longrightarrow a_j \wedge \neg (a_i \stackrel{-1}{\longrightarrow} a_j) \wedge \neg (a_i \stackrel{+1}{\longrightarrow} a_j). 
\end{array}$ 
\end{itemize}
\end{definition}
Figure~\ref{fig:inet-ex1} shows the signed interaction graph of the evolution function~(\ref{eq:ex1}). 

The sign of path or cycle extends the definition to a sign product of the arcs belonging to the path. For example, the sign of the cycle $(a_2,a_3,a_2)$ is negative ($-1$) in Figure~\ref{fig:inet-ex1}. Notice that, the signed interaction graph can be deduced directly from the evolution function where the formulas are in disjunctive normal form\footnote{Recall that the disjunctive normal form of a formula is a non redundant disjunction of clauses where each clause is a conjunction of literals.}. 


\subsection{Model of dynamics}
The discrete Boolean dynamics is modeled by a labeled transition systems (LTS), called the \emph{agent based modal transition system} (AMTS). It represents all the possible trajectories according to a mode. The \emph{modalities} is a set of agents defining the agents that can evolve jointly. They label the arcs. For example $s \xevo{\{a_1,a_2\}} s'$ means that the state of $a_1$ and $a_2$ will be updated and at least one of the states of these agents in $s'$ differ in $s$. $(0,0,0) \xevo{\{a_1,a_2\}} s'$ implies that $s'$ belongs to $\{(0,0,1),(0,1,0),(0,1,1) \}$. The different modalities are gathered into a set called a \emph{mode}. 
\begin{definition}
\label{def:amts}
An AMTS, is a labeled transition system, $\tuple{S,W,\evo}$, where $S$ is a set of states, $W\subseteq \partset{A}$ is the \emph{mode}, and $\evo \subseteq S \times W \times S$ is a relation\footnote{A transition $(s,w,s') \in \evo$ is denoted by $s \xevo w s'$.} on states labeled by a \emph{modality} $w \in W$, such that: 
$$\forall w \in W, \forall s_1,s_2 \in S: s_1 \xevo w s_2 \implies s_1[w] \neq s_2[w] \wedge s_1[A\setminus w] = s_2[A \setminus w].$$ 

\noindent
$\mode \mathcal M = W$ denotes the mode of $\mathcal M$.
\end{definition}
Among the possible modes, some of them are preferentially used for the Boolean discrete dynamics study. For sequential/asynchronous mode where the state of one agent only is updated by a transition, $W$ equals the set of singletons, $W=\{\{a\}\}_{a \in A}$. $W=\{A\}$ defines the \emph{parallel} mode where all the agents can evolve together. The set of part of $A$, $W=\partset{A} \setminus \{ \emptyset \}$, corresponds to the \emph{generalized} mode encompassing all the possible modalities. The parallel and sequential modes partition\footnote{ $\bigcup_{w\in W} w= A \land \forall w_1,w_2 \in W: w_1 \cap w_2 = \emptyset$.} $A$ whereas the generalized mode does not unless $A$ is reduced to a single agent. 

The \emph{spectrum} of a mode (Definition~ \ref{def:spectrum}), denoted by $\spectrum W$, abstracts the structure of a mode by collecting the number of modalities with the same cardinalities. For example, the spectrum corresponding to the following mode $\{\{a_1\}, \{a_2\}, \{a_3,a_4\},\{a_5,a_6,a_7\} \}$ is $\{2\bullet 1,1\bullet 2,1\bullet 3\}$ where $k \bullet m$ means that there exists $k$ modalities of cardinality $m$. For $n$ agents, the spectrum corresponding to the sequential mode is $\{n \bullet 1\}$ and $\{1\bullet n\}$ for the parallel mode. 
For the generalized mode the spectrum is $\{ \tbinom{n}{m}\bullet m\}_{1 \leq m \leq n}$. 

\begin{definition}
\label{def:spectrum}
Let $W$ be a mode, let $W_{m} \subseteq W$ be the set of modalities with the same cardinality $m$, \ie $W_m = \{ w \in W \mid m =|w|\}$, the \emph{spectrum of} $W$, $\spectrum W$ is a multiset ($\spectrum W: \llbracket n \rrbracket \to \llbracket n \rrbracket$) such that: 
$$ \spectrum W = \{ |W_m| \bullet m  \}_{ W_m \neq \emptyset,  1 \leq m \leq n}.$$

\end{definition}
The \emph{regular} modes corresponds to modes with a spectrum of the form $k \bullet m$ such that $km=n$, namely a regular mode corresponds to a partition of the agent into modalities with same cardinality. For example, sequential and parallel modes are regular modes.

The dynamics related to a network $N=\tuple{f_A,W}$ is represented by an AMTS modeling the trajectories on states. Then, an AMTS $\mathcal M$ \emph{models} a network if and only if the transitions are computed from the evolution function (Definition~\ref{def:model}) with respect to the mode $W$. However, an AMTS does not necessary define a model. An AMTS is a model if and only if we can deduce the evolution functions from the transitions with respect a mode. Notice that each network has an unique model but the model varies if the mode varies.

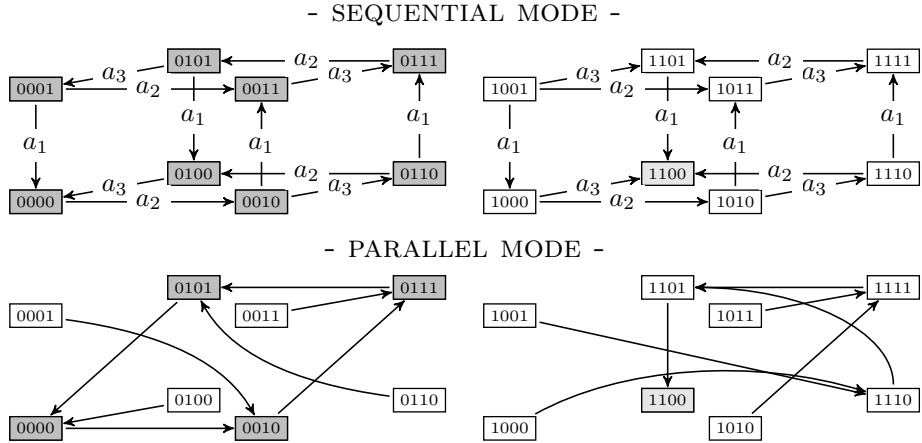
\begin{figure}[ht]
 \begin{center}
 \textsc{- sequential mode -}

 \medskip
\begin{tikzpicture}[xscale=3, yscale=1.5]
\GraphInit[vstyle = normal]
\tikzset{VertexStyle/.style = {
 draw, 
 shape = rectangle, 
 minimum size = 0, 
 line width = 0.5, 
 color = black, 
 fill = white, 
 font=\tiny,
 text = black, 
inner sep = 2.5pt, 
 outer sep = 0.5pt }} 
\Vertex[x=0,y=0]{0000} 
\Vertex[x=0,y=1]{0001} 
\Vertex[x=1,y=0]{0010} 
\Vertex[x=1,y=1]{0011} 
\Vertex[x=.7,y=.25]{0100}
\Vertex[x=.7,y=1.25]{0101}
\Vertex[x=1.7,y=.25]{0110} 
\Vertex[x=1.7,y=1.25]{0111} 
\Vertex[x=2.1,y=0]{1000}
\Vertex[x=2.1,y=1]{1001} 
\Vertex[x=3.1,y=0]{1010} 
\Vertex[x=3.1,y=1]{1011} 
\Vertex[x=2.8,y=0.25]{1100}
\Vertex[x=2.8,y=1.25]{1101} 
\Vertex[x=3.8,y=0.25]{1110}
\Vertex[x=3.8,y=1.25]{1111} 

\tikzset{VertexStyle/.append style={fill=black!10}}
\Vertex[x=2.8,y=0.25]{1100}

\tikzset{VertexStyle/.append style={fill=black!25}}
\Vertex[x=0,y=0]{0000} 
\Vertex[x=0,y=1]{0001} 
\Vertex[x=1,y=0]{0010} 
\Vertex[x=1,y=1]{0011} 
\Vertex[x=.7,y=.25]{0100}
\Vertex[x=.7,y=1.25]{0101}
\Vertex[x=1.7,y=.25]{0110} 
\Vertex[x=1.7,y=1.25]{0111} 
\SetUpEdge[style={post,->}, labelstyle={font=\footnotesize}]

\footnotesize
 \Edge[label={$a_3$}](0100)(0000);
 \Edge[label={$a_3$}](0010)(0110);
 \Edge[label={$a_3$}](1000)(1100);
 \Edge[label={$a_3$}](1010)(1110);
 \Edge[label={$a_3$}](0101)(0001);
 \Edge[label={$a_3$}](0011)(0111);
 \Edge[label={$a_3$}](1001)(1101);
 \Edge[label={$a_3$}](1011)(1111);
 \Edge[label={$a_2$}](0110)(0100);
 \Edge[label={$a_2$}](0111)(0101);
 \Edge[label={$a_2$}](1110)(1100);
 \Edge[label={$a_2$}](1111)(1101);
 \Edge[label={$a_2$}](0000)(0010);
 \Edge[label={$a_2$}](0001)(0011);
 \Edge[label={$a_2$}](1000)(1010);
 \Edge[label={$a_2$}](1001)(1011);
 \Edge[label={$a_1$}](0001)(0000);
 \Edge[label={$a_1$}](0101)(0100);
 \Edge[label={$a_1$}](1001)(1000);
 \Edge[label={$a_1$}](1101)(1100);
 \Edge[label={$a_1$}](0010)(0011);
 \Edge[label={$a_1$}](0110)(0111);
 \Edge[label={$a_1$}](1010)(1011);
 \Edge[label={$a_1$}](1110)(1111);
 \end{tikzpicture}

 \textsc{- parallel mode -}

\medskip
\begin{tikzpicture}[xscale=3, yscale=1.5]
\GraphInit[vstyle = normal]

\tikzset{VertexStyle/.style = {
 draw, 
 shape = rectangle, 
 minimum size = 0, 
 line width = 0.5, 
 color = black, 
 fill = white, 
 font=\tiny, 
 text = black, 
inner sep = 2.5pt, 
 outer sep = 0.5pt }} 
\Vertex[x=0,y=0]{0000} 
\Vertex[x=0,y=1]{0001} 
\Vertex[x=1,y=0]{0010} 
\Vertex[x=1,y=1]{0011} 
\Vertex[x=.7,y=.25]{0100}
\Vertex[x=.7,y=1.25]{0101}
\Vertex[x=1.7,y=.25]{0110} 
\Vertex[x=1.7,y=1.25]{0111} 
\Vertex[x=2.1,y=0]{1000}
\Vertex[x=2.1,y=1]{1001} 
\Vertex[x=3.1,y=0]{1010} 
\Vertex[x=3.1,y=1]{1011} 
\Vertex[x=2.8,y=0.25]{1100}
\Vertex[x=2.8,y=1.25]{1101} 
\Vertex[x=3.8,y=0.25]{1110}
\Vertex[x=3.8,y=1.25]{1111} 
\tikzset{VertexStyle/.append style={fill=black!10}}
\Vertex[x=2.8,y=0.25]{1100}
\tikzset{VertexStyle/.append style={fill=black!25}}
\Vertex[x=0,y=0]{0000} 
\Vertex[x=1,y=0]{0010} 
\Vertex[x=.7,y=1.25]{0101}
\Vertex[x=1.7,y=1.25]{0111} 
\tikzset{EdgeStyle/.style = {post}}
 \Edge(0000)(0010);
 \Edge[style={bend left, ->,post}](0001)(0010);
 \Edge(0010)(0111);
 \Edge(0011)(0111);
 \Edge(0100)(0000);
 \Edge(0101)(0000);
 \Edge[style={bend left,->,post}](0110)(0101);
 \Edge(0111)(0101);
 \Edge[style={bend left=40,->,post}](1000)(1110);
 \Edge(1001)(1110);
 \Edge(1010)(1111);
 \Edge(1011)(1111);
 \Edge(1101)(1100);
 \Edge[style={bend right=45,->,post}](1110)(1101);
 \Edge(1111)(1101);
 \end{tikzpicture}
 \end{center}
 \caption{\small Two models with different modes define the Boolean dynamics of networks with the same the Boolean evolution function (\ref{eq:ex1}): sequential $\{\{a_1\}, \{a_2\}, \{a_3\}, \{a_4\} \}$ and parallel mode $\{\{a_4,a_3,a_2,a_1\}\}$. Stable states are colored in light gray while periodic attractors in gray.}
 \label{fig:model1}
\end{figure}

\begin{definition}
\label{def:model}
Let $N=\tuple{f_A,W}$ be a network for a set $A$ of agents, an AMTS $\mathcal M=\tuple{S,W_{\mathcal M},\evo}$ \emph{models} the dynamics of $N$, denoted $ \mathcal M \models N$, if and only if:
\begin{itemize}
\item the mode is the mode of the network: 

$ W_{\mathcal M} = W$; 
\item the states correspond to all the possible Boolean states of the agents: 

$S =\dom f_A = \Bset^{|A|}$;
\item the transitions are the result of the application of the evolution function with respect to the mode:

$\forall w \in W, \forall s_1,s_2 \in S:s_1 \xevo w s_2 \iff s_2=\tilde f_w(s_1) \wedge s_1 \neq s_2.$
\end{itemize}
\end{definition}
Figure~\ref{fig:model1} depicts models of networks with the evolution function~(\ref{eq:ex1}) by application of the sequential and parallel modes. 

\subsection{Asymptotic dynamics and equilibrium}
 An equilibrium is a specific kind of states encountered at the asymptote of the dynamics and defined as a state infinitely often met by transitions once reached, that is:
\begin{equation*}
\forall s_2 \in S: s_1 \evo^* s_2 \implies s_2 \evo^* s_1.
\end{equation*}
 An \emph{attractor}, $E$, is an equilibrium set where each state (equilibrium) reaches any others (\ie $\forall s \in E: (s \evo^*) = E$). A \emph{steady state} is a specific attractor whose cardinality equals to $1$, $|E|=1$ (\ie $(s \evo^*) = \{s\}$). In the model, an attractor is a terminal strongly connected component whereas a steady state is a sink. 

\section{Isomorphic models}
\label{sec:isomorphic-models} 
\begin{figure}[ht]
 \begin{center}
\begin{tabular}{c c}
\begin{tikzpicture}[xscale=3, yscale=2.5]
\GraphInit[vstyle = normal]
\tikzset{VertexStyle/.style = {
 draw, 
 shape = rectangle, 
 minimum size = 0, 
 line width = 0.5, 
 color = black, 
 fill = white, 
 font=\tiny,
 text = black, 
inner sep = 2.5pt, 
 outer sep = 0.5pt }} 
\Vertex[x=0,y=0]{000} 
\Vertex[x=0,y=1]{001} 
\Vertex[x=1,y=0]{010} 
\Vertex[x=1,y=1]{011} 
\Vertex[x=.7,y=.25]{100}
\Vertex[x=.7,y=1.25]{101}
\Vertex[x=1.7,y=.25]{110} 
\Vertex[x=1.7,y=1.25]{111} 
\SetUpEdge[style={post,->}, labelstyle={font=\footnotesize}]
 \Edge[label={$a_1$}](001)(000); %
 \Edge[label={$a_1$}](010)(011); %
 \Edge[label={$a_1$}](100)(101); %
 \Edge[label={$a_1$}](110)(111); %
 \Edge[label={$a_2$}](010)(000); %
\Edge[label={$a_2$}](011)(001); %
 \Edge[label={$a_2$}](101)(111); %
\Edge[label={$a_2$}](110)(100); %
 \Edge[label={$a_3$}](011)(111); %
 \Edge[label={$a_3$}](100)(000); %
 \Edge[label={$a_3$}](101)(001); %
 \Edge[label={$a_3$}](110)(010); %
 \end{tikzpicture}
& 
\begin{tikzpicture}[xscale=3, yscale=2.5]
\GraphInit[vstyle = normal]
\tikzset{VertexStyle/.style = {
 draw, 
 shape = rectangle, 
 minimum size = 0, 
 line width = 0.5, 
 color = black, 
 fill = white, 
 font=\tiny,
 text = black, 
inner sep = 2.5pt, 
 outer sep = 0.5pt }} 

\Vertex[x=0,y=0]{000} 
\Vertex[x=0,y=1]{110} 
\Vertex[x=1,y=0]{011} 
\Vertex[x=1,y=1]{111} 
\Vertex[x=.7,y=.25]{100}
\Vertex[x=.7,y=1.25]{010}
\Vertex[x=1.7,y=.25]{101} 
\Vertex[x=1.7,y=1.25]{001} 
\SetUpEdge[style={post,->}, labelstyle={font=\footnotesize}]
\footnotesize
 \Edge[label={$a_3,a_2$}](110)(000); %
 \Edge[label={$a_3$}](011)(111); %
 \Edge[label={$a_3,a_2$}](100)(010); %
 \Edge[label={$a_3$}](101)(001); %
 \Edge[label={$a_2,a_1$}](011)(000); 
\Edge[label={$a_1$}](111)(110); 
 \Edge[label={$a_2,a_1$}](010)(001); 
\Edge[label={$a_1$}](101)(100); 
 \Edge[label={$a_3,a_2$}](111)(001); %
 \Edge[label={$a_3$}](100)(000); %
 \Edge[label={$a_3$}](010)(110); %
 \Edge[label={$a_3,a_2$}](101)(011); %
 \end{tikzpicture}
\\
\end{tabular}
\end{center}
\caption{\small 
On the left is represented the sequential model of the evolution function $f_{\{a_3,a_2,a_1\}} = (a_1 \land a_2, a_1 \land a_3, a_2 \lor a_3) $.
On the right, is represented the permuted models by the Boolean permutation $(001 \,110 \, 0101 \, 010 \, 011 \, 111)$ 
 where the transition labels indicate the updated agents but does not refer to a mode. From states $100$ and $101$, we deduce that the presumed mode should include $a_3$ in two distinct modalities. This however conflicts with the updates of states $110$ and $111$.}
\label{fig:nomodel}
\end{figure}

Basically, two models $\mathcal M= \tuple{S,W,\evo}$ and $\mathcal M'= \tuple{S',W',\evo'}$ are isomorphic, $\mathcal M \simeq \mathcal M'$, if and only if there exists a bijection $\varphi:S \to S'$ preserving the transitions:
$$ \forall s_1,s_2 \in S: s_1 \evo s_2 \iff \varphi(s_1) \evo' \varphi(s_2).$$
 Hence the transitions in two isomorphic models may differ on states but preserve the trajectories. In particular, the equilibria of isomorphic models are structurally identical. 
Figure~\ref{fig:iso} shows two isomorphic models for networks with sequential mode where the states of a model correspond to the element-wise negation of the other. 

 However, the new transition relation obtained by application of an isomorphism on states may not produce a model because no mode could be found for the new transition relation (AMTS). In other words, the models are not closed by isomorphism as it is illustrated in Figure~\ref{fig:nomodel}. Thus, we need to check whether a mode exists for an AMTS isomorphic a model to ensure that the resulting AMTS is also a model. 

In this article, we study the isomorphisms preserving the mode. Notice that they implicitly insure the model closure by isomorphism since any isomorphic model has the same mode (\ie $\mode \mathcal M = \mode \mathcal M^\varphi$). 
The set of isomorphisms preserving a particular mode forms a subgroup of the group of isomorphisms (Proposition~\ref{prop:subgroup}).

\begin{proposition} 
\label{prop:subgroup}
The set of isomorphisms preserving a spectrum or a mode forms a group.
\end{proposition}

\begin{figure}[ht]
  \begin{center}
\begin{tabular}{c c}
$ f_{\{a_3,a_2,a_1\}} = \left \{ \begin{array}{l}
f_{a_1} = a_3 \lor a_2\\ f_{a_2} = a_1 \land a_3 \\ f_{a_3} = a_1 \land \lnot a_2
\end{array}\right.$
&
$ f_{\{a_3,a_2,a_1\}} = \left \{ \begin{array}{l}
f_{a_1} = a_3 \land a_2\\ f_{a_2} = a_1 \lor a_3 \\ f_{a_3} = a_1 \lor \lnot a_2
\end{array}\right.$
\\
\\
\begin{tikzpicture}[xscale=3, yscale=2]
\GraphInit[vstyle = normal]
\tikzset{VertexStyle/.style = {
 draw, 
 shape = rectangle, 
 minimum size = 0, 
 line width = 0.5, 
 color = black, 
 fill = white, 
 font=\tiny,
 text = black, 
inner sep = 2.5pt, 
 outer sep = 0.5pt }} 
\Vertex[x=0,y=0]{000} 
\Vertex[x=0,y=1]{001} 
\Vertex[x=1,y=0]{010} 
\Vertex[x=1,y=1]{011} 
\Vertex[x=.7,y=.25]{100}
\Vertex[x=.7,y=1.25]{101}
\Vertex[x=1.7,y=.25]{110} 
\Vertex[x=1.7,y=1.25]{111} 
\tikzset{VertexStyle/.append style={fill=black!10}}
\Vertex[x=0,y=0]{000}
\tikzset{VertexStyle/.append style={fill=black!70, text=white}}
\SetUpEdge[style={post,->}, labelstyle={font=\footnotesize}]
\footnotesize
 \Edge[label={$a_1$}](001)(000); %
 \Edge[label={$a_1$}](010)(011); %
 \Edge[label={$a_1$}](100)(101); %
 \Edge[label={$a_1$}](110)(111); %
 \Edge[label={$a_2$}](010)(000); %
\Edge[label={$a_2$}](011)(001); %
 \Edge[label={$a_2$}](101)(111); %
\Edge[label={$a_2$}](110)(100); %
 \Edge[label={$a_3$}](001)(101); %
 \Edge[label={$a_3$}](100)(000); %
 \Edge[label={$a_3$}](110)(010); %
 \Edge[label={$a_3$}](111)(011); %
 \end{tikzpicture}
& 
\begin{tikzpicture}[xscale=3, yscale=2]
\GraphInit[vstyle = normal]
\tikzset{VertexStyle/.style = {
 draw, 
 shape = rectangle, 
 minimum size = 0, 
 line width = 0.5, 
 color = black, 
 fill = white, 
 font=\tiny,
 text = black, 
inner sep = 2.5pt, 
 outer sep = 0.5pt }} 
\Vertex[x=0,y=0]{111} 
\Vertex[x=0,y=1]{110} 
\Vertex[x=1,y=0]{101} 
\Vertex[x=1,y=1]{100} 
\Vertex[x=.7,y=.25]{011}
\Vertex[x=.7,y=1.25]{010}
\Vertex[x=1.7,y=.25]{001} 
\Vertex[x=1.7,y=1.25]{000} 
\tikzset{VertexStyle/.append style={fill=black!10}}
\Vertex[x=0,y=0]{111}
\tikzset{VertexStyle/.append style={fill=black!70, text=white}}
\SetUpEdge[style={post,->}, labelstyle={font=\footnotesize}]
\footnotesize
 \Edge[label={$a_1$}](001)(000); %
 \Edge[label={$a_1$}](011)(010); %
 \Edge[label={$a_1$}](101)(100); %
 \Edge[label={$a_1$}](110)(111); %
 \Edge[label={$a_2$}](001)(011); %
\Edge[label={$a_2$}](010)(000); %
 \Edge[label={$a_2$}](100)(110); %
\Edge[label={$a_2$}](101)(111); %
 \Edge[label={$a_3$}](000)(100); %
 \Edge[label={$a_3$}](001)(101); %
 \Edge[label={$a_3$}](011)(111); %
 \Edge[label={$a_3$}](110)(010); %
 \end{tikzpicture}
\\
\\
\multicolumn{2}{c}{
 \begin{tikzpicture}[scale=2, node distance=2cm]
 \SetVertexNormal[Shape = circle, LineWidth=0.5pt]
\SetUpEdge[style={post}]
\tikzset{LabelStyle/.style = {sloped}}
\SetVertexMath

\Vertex[ x=0,y=0]{a_1} 
\Vertex[x=2,y=0]{a_2} 
\Vertex[x=1,y=1]{a_3} 
 \Edge[ label= + ,labelstyle=above](a_2)(a_1)
 \Edge[ label= + ,labelstyle=above, style={bend right, post}](a_3)(a_1)
 \Edge[ label= + ,labelstyle=below, style={bend right, post}](a_1)(a_2)
 \Edge[ label= + ,labelstyle=below](a_3)(a_2)
 \Edge[ label= + ,labelstyle=below](a_1)(a_3)
 \Edge[ label= - ,labelstyle=above, style={bend right, post}](a_2)(a_3)
\end{tikzpicture}
} 
\end{tabular}
\end{center}
\caption{\small The isomorphism associates each state with its element-wise complement. The interaction networks are identical.}
\label{fig:iso}
\end{figure}

\subsection{Notations and definitions on group}
\label{sec:definition}
In this section, we recall some definitions related to group of isomorphisms.  

\paragraph{Groups of  permutations} We use three kinds of  isomorphisms: the group of \emph{Boolean permutations} on $\Bset^l, l \in \Nset$ denoted $\gsym{\Bset^l}$, the group of integer permutations on $\llbracket l \rrbracket = \{1, \ldots, l\}, l \in \Nset$ denoted $S_l$ and the group of \emph{signed permutations} of rank $l, l \in \Nset$ denoted $BC_l$.  $e_G$ is the identity of the group $G$ (\ie $\forall g \in G: e_G \, g = g \, e_G = g$).  Throughout the article we adopt the following notations for the isomorphisms:
 $\beta$ stands for a Boolean permutation, $\pi$ for a permutation, $\sigma$ for a signed permutation and $\varphi$ for an isomorphism preserving the mode.

The action of a permutation $\varphi$ on elements uses the classical exponent notation (\ie $s^\varphi = \varphi(s)$). Its extension on sets corresponds to the action on each element (\ie $ S^\varphi = \{s_1^\varphi, \ldots, s_n^\varphi \}$).
Notice  that the groups are defined up to an isomorphism on groups. In particular, the following equivalence is used for the proofs: $\forall l \in \Nset, \gsym{\Bset^l} \simeq S_{2^l}$.

\paragraph{Group of signed permutation}  $BC_l$ is the group of signed permutations of rank $l$ defined on $I=\{-l,\ldots,-1,1,\ldots,l\}$ such that: 
$$\sigma({-i}) = - \sigma({i}), i \in I,$$ 
(\eg $1 \mapsto -2, -1 \mapsto 2$). The action of a signed permutation $\sigma$ on $\Bset^l$ is explained as permuting the sequence of $0'$s and $1'$s and then taking the negation at some positions. Given a signed permutation $\sigma$,
we define $\pi_\sigma$ its non-signed permutation (\ie $\pi_\sigma(i)=|\sigma(i)|, i \in \llbracket l \rrbracket$); $\sigma$ acts on $\Bset^l$ as follows: 
\begin{equation*}
b \in \Bset^l, b^{\sigma} = c \;\text{where}\; c_i = \left \{ 
\begin{array}{l @{\; \text{iff }\;} l}
b[{\pi_\sigma^{-1}(i)}] & \sigma^{-1}(i) > 0 \\
\lnot{b[{\pi_\sigma^{-1}(i)}]} & \sigma^{-1}(i) < 0
\end{array}
\right.
\end{equation*}
For example, the action of $\sigma = \{ 1 \mapsto -2, 2 \mapsto 1, -1 \mapsto 2 , -2 \mapsto -1 \}$ on $(1,0)$ leads to $(1,0)^{\sigma} = (1,1)$. 
A convenient representation of a signed permutation is a pair $\sigma= (\pi_\sigma,p), \pi \in S_{n}, p \in \Bset^n$ where 
$p(i)=1$ if the $i^\text{ith}$ element is negated ($\sigma(i)<0$) and $0$ otherwise. Hence, the action of $\sigma=(\pi_\sigma,p)$ is then defined as follows\footnote{Recall that $x \lxor 1 = \lnot x$ and $x \lxor 0 = x$, $x \in \Bset$.}: 
\begin{eqnarray*}
 c &=& (b \lxor p)^{\pi_\sigma} \quad \text{where} \\
 c_i &=& b[{\pi_\sigma^{-1}(i)}] \lxor p[\pi_\sigma^{-1}(i)], i \in \llbracket l \rrbracket. 
\end{eqnarray*}
 For example, the signed permutation inverting the elements of a vector $b\in \Bset^3$ (Figure~\ref{fig:iso}) is defined as $((1,1,1),e_{S_3})$. 

\paragraph{Representation of permutation} A permutation will be represented using the classical factorization in disjoint cycles where each cycle, written $(c_1\,\ldots\,c_k)$, is a permutation $\{c_1 \mapsto c_2, \ldots, c_k \mapsto c_1 \}$. For instance, the cycle $(1\,2)(3\,4)$ corresponds to the mapping $\{ 1 \mapsto 2, 2 \mapsto 1, 3 \mapsto 4 ,4 \mapsto 3, 5 \mapsto 5\}$. By convention, elements mapped to themselves are omitted like $5$ in the example. 

\subsection{Group of isomorphisms preserving the mode}

Preserving a mode by an isomorphism depends on both a model and a mode. For instance, by considering the model of the network with the identity as evolution function, any mode is preserved by any isomorphism due to the absence of transition. However, the isomorphisms preserving a mode for this particular model may not preserve it for another model. We are seeking here for a more general notion that depends on the mode only, namely: the group of isomorphisms preserving the mode whatever the model. 

To characterize this group,  we first specifically focus on those preserving the regular mode of length $m$ for $n$ agents, $SRM^k_m, k= \frac{n}{m}$  because they constitute the basic bricks for characterizing the group preserving the mode partitioning the agents. For example, the groups preserving the sequential and parallel modes respectively correspond to $SRM_1^n$ and $SRM_n^1$. The characterization of such groups follows a general scheme introduced for the group preserving the sequential mode. 

Informally the scheme is as follows: we define the group of automorphisms of a graph collecting all the models with the same set of states and the same mode. As the automorphism preserve the structure of the graph, their action on  a model corresponding to a sub-graph necessary maps in another model (sub-graph). Conversely, an isomorphism which is not an automorphism of this graph does not preserve the mode for some models. Thus, the group of the isomorphisms preserving the mode is the group of automorphisms of this graph.

\paragraph{Group of isomorphisms preserving the sequential mode}  For the sequential mode, two states are related by a transition if and only if they differ in one position only. Thus any pair of vertices differing in one position only is an arc of this graph. Hence  collecting all the models with a sequential mode generates an $n$-dimensional hypercube $Q_n$ \cite{Harary1988} whose definition is based on the Hamming distance, $\operatorname{hd}:\Bset^n \times \Bset^n \to \Nset$: $\operatorname{hd}(b_1,b_2)=\sum_{i=1}^n( b_1(i) \lxor b_2(i))$. Formally a $n$-dimensional hypercube $Q_n$ is a graph such that: 

\medskip
$ V(Q_n)=\Bset^n \;\text{and}\; E(Q_n)=\{(b_1,b_2) \mid \operatorname{hd}(b_1,b_2) = 1 \}.$

\medskip
The group of automorphisms of the hypercube is also known to be isomorphic to the group of signed permutations $BC_n$~\cite{Chen1993}. Hence, the order of the sub-group preserving the sequential mode for $n$ agents is $2^{n}n!$.
\begin{lemma} 
\label{lem:sequential-group}
The group of the signed permutation of rank $n$ is isomorphic to the group preserving the asynchronism: $\quad \aut(Q_n)\simeq BC_n \simeq SRM_1^n.$
\end{lemma}

\paragraph{Group of isomorphisms preserving the regular mode}
We generalize the afore result to the groups preserving a regular mode. Their characterization is based on the decomposition of graphs in a product of prime factors\footnote{A graph is prime if it is decomposed as a product of trivial graphs only, \ie $G$ is prime if and only if: $G = G_1 \Box G_2$ implies that $G_1=G$ or $G_2=G$.}.
 The reader may refer to \cite{Imrich2008, Hammack2011, Sabidussi1959} for a coverage of the topics.

The Cartesian product of two graphs, $G_1 \Box G_2$, is a binary operation on graphs defined as follows:

\medskip
\noindent
$V(G_1 \Box G_2) = V(G_1) \times V(G_2)$ and,

\noindent
$\begin{array}[t]{l l} 
E(G_1 \Box G_2)= &\{ ( v_1v_2,v_1'v_2) \mid (v_1,v_1') \in E(G_1) \} \cup \\		
 					 &\{ ( v_1v_2,v_1v_2') \mid (v_2,v_2') \in E(G_2) \}.
\end{array}$

\medskip

First, we consider the graph union of models with a regular mode $W$, called the \emph{complete modal graph} of $W$, $KM_W$, and defined as follows:

\medskip
\noindent
$V(KM_W)= \Bset^n$ and,

\noindent
$E(KM_W)=\{ (b_1,b_2) \mid b_1[w] \neq b_2[w] \wedge b_1[A \setminus \{w\}]= b_2[A \setminus \{w\}], w \in W \}$.

\medskip
A complete modal graph of a regular mode $W$ whose length is $m$ isomorphic to the Cartesian product of $\frac{n}{m}$ complete graphs $K_{2^m}$ (Proposition~\ref{prop:srm}). 

\begin{proposition}
\label{prop:srm}
Let $W$ be a regular mode of length $m$, the complete modal graph $KM_W$ is isomorphic to $K_{2^m}^{\Box \frac{n}{m}}$.
\end{proposition}
 Hence, the group of isomorphisms preserving a regular mode of length $m$ is isomorphic to $\aut( (K_{2^m})^{\Box \frac{n}{m}})$, as $KM_W \simeq (K_{2^m})^{\Box \frac{n}{m}}$ for all regular mode $W$ of length $m$. Lemma~\ref{lem:srm} characterizes the group of automorphisms of complete graph product in term of permutations. The order\footnote{The order of a wreath product between groups of symmetry $S_l \wr S_k$ is $(l!)^k k!$.} of this group is $\left(2^m!\right)^{k} k!$.

\begin{lemma} 
\label{lem:srm}
The wreath product of $\gsym{\Bset^m} \wr S_k$ is isomorphic to the group preserving the regular mode of length $m$:
$\quad SRM_m^k \simeq \aut( (K_{2^m})^{\Box \frac{n}{m}}) \simeq \gsym{\Bset^m} \wr S_k.$
\end{lemma}

The characterization of the group preserving regularity as a wreath product (Lemma~\ref{lem:srm}) leads to  a valuable representation of isomorphism preserving the regular mode for investigating the properties related to model isomorphism (Section~\ref{sec:evo-fun}).  
The wreath product is an operation on groups acting on partitioned set~\cite{Cameron1999}. An isomorphism of $SRM^k_m$   is represented as a pair $\varphi=(\beta, \pi)$ where $\beta: \llbracket k \rrbracket \to \gsym{\Bset^m} $ is a function assigning a Boolean permutation of $\gsym{\Bset^m}$ to each $i \in \llbracket k \rrbracket$ such that $\beta_i$ acts on $b[w_i]$; and $\pi: \llbracket k \rrbracket \to \llbracket k \rrbracket$ is an integral  permutation  leading to exchange the position of the modalities arbitrary indexed that concretely leads to an ``agent renaming'' by block of modalities. Thus, the action is in twofold: first inside each modality $w_i \in W$, by a  Boolean permutations of $\gsym{\Bset^m}$ acting on the sub-vector $b[w_i]$, and second among modalities by exchanging the location of the sub-vectors. The last operation can be assimilated to an ``agent renaming'' while preserving the block structure of modalities. Accordingly, the action of Boolean vectors is defined as:
\begin{equation}
c= b^\varphi \; \text{where} \; c=\left( b[w_{\pi^{-1}(1)}]^{\beta_{\pi^{-1}(1)}}, \ldots, b[w_{\pi^{-1}(k)}]^{\beta_{\pi^{-1}(k)}} \right), b,c \in \Bset^n.
\label{eq:action}
\end{equation}
For example, given the mode $\{\overbracket{\{a_4,a_3\}}^{1},\overbracket{\{a_2,a_1\}}^{2}\}$ the action of:
$$\varphi=(\beta_1=(00\; 11),\beta_2= (01\;10),\pi = (1 \, 2))$$ on the following Boolean vectors\footnote{Recall that the order of states in a Boolean vector is $(a_4,a_3,a_2,a_1)$ by convention.} is:
 $$0010^\varphi = 0111, 1110^\varphi=0100, 1001^\varphi=1010, 0111^\varphi=1101.$$

 Lemma~\ref{lem:srm} generalizes Lemma~\ref{lem:sequential-group}. Indeed, the group $\gsym{\Bset} \wr S_{n} $ is isomorphic to the group of the signed permutations $BC_n$, (\ie $\gsym{\Bset} \wr S_n \simeq S_{2} \wr S_n \simeq BC_n$)~\cite{Harary2000}.
Notice that the group of isomorphism preserving the parallel mode $SRM^1_n$ is obviously isomorphic to the group of all Boolean permutations $\gsym{\Bset^n}$ ($\gsym{\Bset^n} \wr S_1 = \gsym{\Bset^n}$).

\paragraph{Group of isomorphisms preserving partition}  $SP_{\spectrum W}$  is the group preserving the partition $W$ of $A$. As mode $W$ can be defined as an union of regular sub-modes $W=\bigcup_{i=1}^l W_i$ such that $\spectrum W_i = k_i \bullet m_i$,  the spectrum is of the form $\spectrum W = \{ k_i \bullet m_i \}_{1 \leq i \leq l}$.  $SP_{\spectrum W}$ is proved to be the product of isomorphisms preserving regular mode (Theorem~\ref{the:sm}). 
\begin{theorem} 
\label{the:sm}
The set of isomorphisms on models preserving a mode partitioning the agents, the spectrum of which is $ \gamma= \{k_i \bullet m_i\}_{1 \leq i \leq l}$, is the group defined as the Cartesian product of their regular modes:
$$\quad SP_\gamma = \bigtimes_{i=1}^l SRM^{k_i}_{m_i}, SRM^{k_i}_{m_i} \simeq \gsym{\Bset^{m_i}} \wr S_{k_i}.$$
\end{theorem}
Thus, the order of this group is $\prod_{i=1}^l\left(2^{m_i}!\right)^{k_i} k_i!$.
The isomorphisms of  $SP_{\spectrum W}$ are also represented by a pair $(\beta,\pi)$ respectively referring to a collection of Boolean permutations and a permutation on modalities. However, $\pi$ is constrained by the fact that only the permutations of modalities with the same cardinalities are allowed. Thereby the action of an isomorphism $\varphi=(\beta,\pi) \in SP_{\spectrum W}$ is also defined by (\ref{eq:action}). By extension, the  action on transitions  is defined as:
\begin{equation}
\label{eq:equiv-iso}
\forall w_i \in W, \forall s_1,s_2 \in \Bset^n:(s_1 \xevo{w_i} s_2)^\varphi= s_1^\varphi \xevo{w_{\pi(i)}} s_2^\varphi.
\end{equation}



\section{Equivalence on networks}
\label{sec:evo-fun}
The equivalence formalizes the analogy on the behaviors of networks with the same mode but with different evolution functions.  Two networks with identical modes are \emph{dynamically equivalent} ($N \sim N'$) if and only if their model is isomorphic with respect to $SP$ group (Definition~\ref{def:equivnet}).
\begin{definition}
\label{def:equivnet}
$$N \sim N' \eqdef \mode N=\mode N' \wedge (\exists \varphi \in SP_{\spectrum \mode N}: \mathcal M \models N 
\wedge \mathcal M^\varphi  \models N').$$
\end{definition}

 Hence, there exists at most $|SP_{\spectrum \mode N}|$ equivalent networks to network $N$. 
The following property deduced from (\ref{eq:equiv-iso}) defines the relation on the functions of two equivalent networks: if $\tuple{f_A,W} \sim \tuple{f'_A,W}$ with respect to $\varphi=(\beta,\pi) \in SP_{\spectrum W} $ we have:
\begin{equation}
\label{eq:equiv-fun}
\forall w_i \in W, \forall s \in \Bset^n: f'_{w_i}(s) = f_{w_{\pi^{-1}(i)}}(s^{\varphi^{-1}})^\varphi.
\end{equation}

\begin{figure}[ht]
 \begin{center}
\begin{tabular}{c c}
$ f_{\{a_2,a_1\}} = \left \{ \begin{array}{l}
f_{a_1} = a_2 \\ f_{a_2} = \lnot{a_1}
\end{array}\right.$
&
$ f'_{\{a_2,a_1\}} = \left \{ \begin{array}{l}
f'_{a_1} = \lnot{a_1} \\ f'_{a_2} = a_1 \lxor a_2
\end{array}\right.$
\\
\\
\multicolumn{2}{c}{\small Interaction graph}
\\
 \begin{tikzpicture}[scale=1, node distance=2cm, baseline]
 \SetVertexNormal[Shape = circle, LineWidth=0.5pt]
 \Vertices[Math, x=0, y=0, dir=\EA]{a_1,a_2};
 \Edge[ label= - ,labelstyle=above, style={post, bend left}](a_1)(a_2)
 \Edge[ label= + ,labelstyle=below, style={post,bend left}](a_2)(a_1)
\end{tikzpicture}
&
 \begin{tikzpicture}[scale=1, node distance=2cm, baseline]
 \SetVertexNormal[Shape = circle, LineWidth=0.5pt]
 \Vertices[Math, x=0, y=0, dir=\EA]{a_1,a_2};
 \Edge[ label= $\pm$ ,labelstyle=above, style={post}](a_1)(a_2)
 \Loop[ label= - ,labelstyle=left, style={post}, dir=WE, dist=1.25cm](a_1)
\end{tikzpicture}
\\
\multicolumn{2}{c}{\small Permutation: $(00 \; 11 \; 01 \; 10)$}
\\
\\
\multicolumn{2}{c}{\small Parallel mode: $\{ \{a_2,a_1 \}\}$}
\\
\begin{tikzpicture}[xscale=3, yscale=1.5]
\GraphInit[vstyle = normal]
\tikzset{VertexStyle/.style = {
 draw, 
 shape = rectangle, 
 minimum size = 0, 
 line width = 0.5, 
 color = black, 
 fill = white, 
 font=\tiny,
 text = black, 
inner sep = 2.5pt, 
 outer sep = 0.5pt }} 
\Vertex[x=0,y=0]{00} 
\Vertex[x=0,y=1]{01} 
\Vertex[x=1,y=0]{10} 
\Vertex[x=1,y=1]{11} 
\SetUpEdge[style={post,->}, labelstyle={font=\footnotesize}]
\footnotesize
 \Edge[label={$a_2,a_1$}](10)(11); 
 \Edge[label={$a_2,a_1$}](01)(00); 
 \Edge[label={$a_2,a_1$}](00)(10); 
 \Edge[label={$a_2,a_1$}](11)(01); 
 \end{tikzpicture}
& 
\begin{tikzpicture}[xscale=3, yscale=1.5]
\GraphInit[vstyle = normal]
\tikzset{VertexStyle/.style = {
 draw, 
 shape = rectangle, 
 minimum size = 0, 
 line width = 0.5, 
 color = black, 
 fill = white, 
 font=\tiny,
 text = black, 
inner sep = 2.5pt, 
 outer sep = 0.5pt }} 
\Vertex[x=0,y=0]{11} 
\Vertex[x=0,y=1]{10} 
\Vertex[x=1,y=0]{00} 
\Vertex[x=1,y=1]{01} 
\SetUpEdge[style={post,->}, labelstyle={font=\footnotesize}]
\footnotesize
 \Edge[label={$a_2,a_1$}](11)(00); 
 \Edge[label={$a_2,a_1$}](01)(10); 
 \Edge[label={$a_2,a_1$}](00)(01); 
 \Edge[label={$a_2,a_1$}](10)(11); 
 \end{tikzpicture}
\\
\\
\multicolumn{2}{c}{\small Sequential mode: $\{\{a_2\},\{a_1\}\}$ }
\\
\begin{tikzpicture}[xscale=3, yscale=1.5]
\GraphInit[vstyle = normal]
\tikzset{VertexStyle/.style = {
 draw, 
 shape = rectangle, 
 minimum size = 0, 
 line width = 0.5, 
 color = black, 
 fill = white, 
 font=\tiny,
 text = black, 
inner sep = 2.5pt, 
 outer sep = 0.5pt }} 
\Vertex[x=0,y=0]{00} 
\Vertex[x=0,y=1]{01} 
\Vertex[x=1,y=0]{10} 
\Vertex[x=1,y=1]{11} 
\SetUpEdge[style={post,->}, labelstyle={font=\footnotesize}]
\footnotesize
 \Edge[label={$a_1$}](10)(11); 
 \Edge[label={$a_1$}](01)(00); 
 \Edge[label={$a_2$}](00)(10); 
 \Edge[label={$a_2$}](11)(01); 
 \end{tikzpicture}
& 
\begin{tikzpicture}[xscale=3, yscale=1.5]
\GraphInit[vstyle = normal]
\tikzset{VertexStyle/.style = {
 draw, 
 shape = rectangle, 
 minimum size = 0, 
 line width = 0.5, 
 color = black, 
 fill = white, 
 font=\tiny,
 text = black, 
inner sep = 2.5pt, 
 outer sep = 0.5pt }} 
\Vertex[x=0,y=0]{00} 
\Vertex[x=0,y=1]{01} 
\Vertex[x=1,y=0]{10} 
\Vertex[x=1,y=1]{11} 
\SetUpEdge[style={post,->,bend left}, labelstyle={font=\footnotesize}]
\footnotesize
 \Edge[label={$a_1$}](10)(11); 
 \Edge[label={$a_1$}](11)(10); 
\Edge[label={$a_1$}](00)(01); 
\Edge[label={$a_1$}](01)(00); 
 \Edge[label={$a_2$}](11)(01); 
 \Edge[label={$a_2$}](01)(11); 
 \end{tikzpicture}
\end{tabular}
\end{center}
\caption{Sensitivity of the network equivalence to the mode.}
\label{fig:mode-depend}
\end{figure}
In this section, we study the structural invariance property of equivalent networks and conditions for preserving the equivalence across different modes. The structural invariance property aims at characterizing a structure abstracting a network such that these structures are isomorphic for equivalent networks. Preserving the equivalence while changing the mode implies to identify properties insuring that an equivalence found for a mode is also preserved for another.
 In fact, the equivalence may not be preserved by mode variation in general. Figure~\ref{fig:mode-depend} shows an example where the models are isomorphic for the parallel mode and not for the sequential one. Moreover, it also shows that the interaction graph is not an invariant structure. These two properties actually depend on more complex relations on modes. 

\subsection{Structural invariance} Although an interaction graph is not an invariant structure (\cf Figure~\ref{fig:mode-depend}), another structure derived from it, called the \emph{interaction modal graph} (IMG), proves to be invariant for equivalent networks (Lemma~\ref{lem:invariant}). The IMG is defined as the unsigned quotient interaction graph induced by a mode~(Definition~\ref{def:img}) where vertices are modalities and the edge set represents the quotient relation of the interactions with regard to a mode. 

\begin{definition}
\label{def:img}
Let $G_f=\tuple{A, \stackrel{\star}{\longrightarrow}}$ be an interaction graph of $f$ and $W$ a mode partitioning $A$, the \emph{interaction modal graph} (IMG) of $f$ for $W$, $G_f/W$, is the unsigned quotient interaction digraph induced by $W$ defined as:
\begin{itemize}
\item $V(G_f/W)=W$;
\item $E(G_f/W)=\{(w_i,w_j)\mid w_i \neq w_j \wedge (\exists a_i \in w_i, \exists a_j \in w_j: a_i \stackrel{\star}{\longrightarrow} a_j )\}.$
\end{itemize}
\end{definition}

\begin{figure}[p]
\begin{center}
\textsc{model \& interaction graph}

of $f_{\{a_4,a_3,a_2,a_1\}} = (a_3,a_4, a_1 \land a_3 \lor a_4, a_4) \; \text{with} \; W=\{\{a_4,a_3\},\{a_2,a_1\}\}.$

\begin{tikzpicture}[baseline,xscale=2.2,yscale=1.5]
\GraphInit[vstyle = normal]
\tikzset{VertexStyle/.style = {
 draw, shape = rectangle,line width = 0.5,inner sep = 2.5pt,outer sep = 0.5pt,minimum size = 0,color = black,fill=white,font=\tiny,text = black }}
\SetUpEdge[style={bend right, post,->}, labelstyle={font=\footnotesize, sloped}]
\definecolor{col0000}{rgb}{0.85, 0.85, 0.85} \tikzset{VertexStyle/.append style={fill=col0000}}
\Vertex[x=2.00003,y=-3.16669]{0000}
\tikzset{VertexStyle/.append style={fill=white}}
\Vertex[x=2.00003,y=-2.16669]{0001}
\Vertex[x=3.00003,y=-3.16669]{0010}
\Vertex[x=3.00003,y=-2.16669]{0011}
\definecolor{col0100}{rgb}{0.75,0.75,0.75} \tikzset{VertexStyle/.append style={fill=col0100}}
\Vertex[x=0.333343,y=-3.33334]{0100}
\tikzset{VertexStyle/.append style={fill=white}}
\Vertex[x=0.333343,y=-2.33334]{0101}
\definecolor{col0110}{rgb}{0.75,0.75,0.75} \tikzset{VertexStyle/.append style={fill=col0110}}
\Vertex[x=0.333343,y=-1.33334]{0110}
\tikzset{VertexStyle/.append style={fill=white}}
\definecolor{col0111}{rgb}{0.75,0.75,0.75} \tikzset{VertexStyle/.append style={fill=col0111}}
\Vertex[x=0.333343,y=-0.333343]{0111}
\tikzset{VertexStyle/.append style={fill=white}}
\definecolor{col1000}{rgb}{0.75,0.75,0.75} \tikzset{VertexStyle/.append style={fill=col1000}}
\Vertex[x=1.33334,y=-3.33334]{1000}
\tikzset{VertexStyle/.append style={fill=white}}
\Vertex[x=1.33334,y=-2.33334]{1001}
\definecolor{col1010}{rgb}{0.75,0.75,0.75} \tikzset{VertexStyle/.append style={fill=col1010}}
\Vertex[x=1.33334,y=-1.33334]{1010}
\tikzset{VertexStyle/.append style={fill=white}}
\definecolor{col1011}{rgb}{0.75,0.75,0.75} \tikzset{VertexStyle/.append style={fill=col1011}}
\Vertex[x=1.33334,y=-0.333343]{1011}
\tikzset{VertexStyle/.append style={fill=white}}
\Vertex[x=2.00003,y=-1.33334]{1100}
\Vertex[x=2.00003,y=-0.333343]{1101}
\Vertex[x=3.00003,y=-1.33334]{1110}
\definecolor{col1111}{rgb}{0.85,0.85,0.85} \tikzset{VertexStyle/.append style={fill=col1111}}
\Vertex[x=3.00003,y=-0.333343]{1111}
\tikzset{VertexStyle/.append style={fill=white}}
\Edge[](0001)(0000)
\Edge[](0010)(0000)
\Edge[](0011)(0000)
\Edge[](0100)(1000)
\Edge[](0101)(0110)
\Edge[](0101)(1001)
\Edge[](0110)(0100)
\Edge[](0110)(1010)
\Edge[](0111)(0110)
\Edge[](0111)(1011)
\Edge[](1000)(0100)
\Edge[](1000)(1011)
\Edge[](1001)(0101)
\Edge[](1001)(1011)
\Edge[](1010)(0110)
\Edge[](1010)(1011)
\Edge[](1011)(0111)
\Edge[](1100)(1111)
\Edge[](1101)(1111)
\Edge[](1110)(1111)
\end{tikzpicture} 
\hspace{1cm}
\begin{tikzpicture}[baseline=3cm,scale=1,node distance=1.5cm]
\tikzstyle{VertexStyle}=[draw,shape = circle, line width=1pt,font=\footnotesize]
\SetUpEdge[ lw = 2pt, color = black, style={ post}, labelcolor=white, labeltext = black, labelstyle = {sloped}]
\Vertices*[Math]{circle}{a_4,a_3,a_2,a_1}
\Edge[label=+, style={bend left, post}](a_4)(a_3)
\Edge[label=+](a_4)(a_2)
\Edge[label=+](a_4)(a_1)
\Edge[label=+, style={bend left,post}](a_3)(a_4)
\Edge[label=+](a_3)(a_2)
\Edge[label=+](a_1)(a_2)
\end{tikzpicture}
\\ 

\bigskip
\textsc{patterns of interaction graphs} 

for all equivalent functions.

\begin{tabular}{ccccc}
\begin{tikzpicture}[baseline,scale=0.85,node distance=1cm]
\tikzstyle{VertexStyle}=[draw,shape = circle, line width=1pt,font=\footnotesize]
\SetUpEdge[ lw = 2pt, color = black, style={ post}, labelcolor=white, labeltext = black, labelstyle = {sloped}]
\SetVertexNoLabel
\Vertices*[Math]{circle}{a_4,a_3,a_2,a_1}
\Edge[style={bend left,post}](a_4)(a_3)
\Edge[](a_4)(a_2)
\Edge[](a_4)(a_1)
\Edge[style={bend left,post}](a_3)(a_4)
\Edge[](a_3)(a_2)
\Edge[](a_1)(a_2)
\end{tikzpicture}
&
\begin{tikzpicture}[baseline,scale=0.85,node distance=1cm]
\tikzstyle{VertexStyle}=[draw,shape = circle, line width=1pt,font=\footnotesize]
\SetUpEdge[ lw = 2pt, color = black, style={ post}, labelcolor=white, labeltext = black, labelstyle = {sloped}]
\SetVertexNoLabel
\Vertices*[Math]{circle}{a_4,a_3,a_2,a_1}
\Edge[style={bend left, post}](a_4)(a_3)
\Edge[](a_4)(a_2)
\Edge[](a_4)(a_1)
\Edge[style={bend left, post}](a_3)(a_4)
\Edge[](a_3)(a_2)
\Edge[](a_3)(a_1)
\Loop[dist=2.5em,dir=WE](a_2)
\Edge[style={bend left, post}](a_2)(a_1)
\Edge[style={bend left, post}](a_1)(a_2)
\Loop[dist=2.5em,dir=SO](a_1)
\end{tikzpicture}
&
\begin{tikzpicture}[baseline,scale=0.85,node distance=1cm]
\tikzstyle{VertexStyle}=[draw,shape = circle, line 
width=1pt,font=\footnotesize]
\SetUpEdge[ lw = 2pt, color = black, style={post}, 
labelcolor=white, labeltext = black, labelstyle = {sloped}]
\SetVertexNoLabel
\Vertices*[Math]{circle}{a_4,a_3,a_2,a_1}
\Loop[dist=2.5em,dir=EA](a_4)
\Edge[](a_4)(a_2)
\Edge[](a_4)(a_1)
\Edge[](a_3)(a_4)
\Loop[dist=2.5em,dir=NO](a_3)
\Edge[](a_3)(a_2)
\Edge[](a_1)(a_2)
\end{tikzpicture}
&
\begin{tikzpicture}[baseline,scale=0.85,node distance=1cm]
\tikzstyle{VertexStyle}=[draw,shape = circle, line width=1pt,font=\footnotesize]
\SetUpEdge[ lw = 2pt, color = black, style={post}, labelcolor=white, labeltext = black, labelstyle = {sloped}]
\SetVertexNoLabel
\Vertices*[Math]{circle}{a_4,a_3,a_2,a_1}
\Loop[dist=2.5em,dir=EA](a_4)
\Edge[](a_4)(a_2)
\Edge[](a_4)(a_1)
\Edge[](a_3)(a_4)
\Loop[dist=2.5em,dir=NO](a_3)
\Edge[](a_3)(a_2)
\Edge[](a_3)(a_1)
\Loop[dist=2.5em,dir=WE](a_2)
\Edge[ style={bend left,post}](a_2)(a_1)
\Edge[ style={bend left,post}](a_1)(a_2)
\Loop[dist=2.5em,dir=SO](a_1)
\end{tikzpicture}
&
\begin{tikzpicture}[baseline,scale=0.85,node distance=1cm]
\tikzstyle{VertexStyle}=[draw,shape = circle, line width=1pt,font=\footnotesize]
\SetUpEdge[ lw = 2pt, color = black, style={ post}, labelcolor=white, labeltext = black, labelstyle = {sloped}]
\SetVertexNoLabel
\Vertices*[Math]{circle}{a_4,a_3,a_2,a_1}
\Loop[dist=2.5em,dir=EA](a_4)
\Edge[](a_4)(a_2)
\Edge[](a_4)(a_1)
\Edge[](a_3)(a_4)
\Loop[dist=2.5em,dir=NO](a_3)
\Edge[](a_3)(a_2)
\Edge[](a_3)(a_1)
\Edge[](a_1)(a_2)
\end{tikzpicture}
\\
256 & 128 & 256 &256 & 256
\end{tabular}

\bigskip
\textsc{interaction modal graphs}

of all interaction graphs.

\bigskip
\begin{tabular}{c @{\hspace{2cm}} c}
 \begin{tikzpicture}[scale=1, node distance=2cm] 
 \tikzstyle{VertexStyle}=[draw, line width=1pt,font=\footnotesize]
 \SetUpEdge[ lw = 2pt, color = black, style={post}, labelcolor=white, labeltext = black, labelstyle = {sloped}]
\Vertex[L={$\{a_2,a_1\}$}, x=0,y=0]{x}
\Vertex[L={$\{a_4,a_3\}$}, x=3,y=0]{y}
	\Edge[](x)(y)
\end{tikzpicture}
&
 \begin{tikzpicture}[scale=1, node distance=2cm] 
 \tikzstyle{VertexStyle}=[draw, line width=1pt,font=\footnotesize]
 \SetUpEdge[ lw = 2pt, color = black, style={post}, labelcolor=white, labeltext = black, labelstyle = {sloped}]
\Vertex[L={$\{a_2,a_1\}$}, x=0,y=0]{x}
\Vertex[L={$\{a_4,a_3\}$}, x=3,y=0]{y}
	\Edge[](y)(x)
\end{tikzpicture} \\
576 & 576
\end{tabular}
\end{center}

\caption{Analysis of all the equivalent networks with $f_{\{a_4,a_3,a_2,a_1\}} = (a_3,a_4, a_1 \land a_3 \lor a_4, a_4)$
 for $W=\{\{a_4,a_3\},\{a_2,a_1\}\}$.}
	\label{fig:img}
\end{figure}
Figure~\ref{fig:img} describes the exhaustive analysis of the networks equivalent to the network $\tuple{(a_3,a_4, a_1 \land a_3 \lor a_4, a_4), \{\{a_4,a_3\},\{a_2,a_1\}\}}$. The regular group corresponding to the mode is $SRM^2_2$ whose order is $1152$.
 All the functions of equivalent networks have been inferred from the isomorphic models obtained by application of the isomorphisms of $SRM^2_2$ to the model of the studied function. The interaction graphs of all the functions follow $5$ patterns but signs and locations of agents vary. 
For each pattern, the number of related interaction graphs is reported below. 
They confirm that interaction graphs of functions for equivalent networks are not isomorphic because the patterns are not. Finally, 2 isomorphic interactions modal graphs are derived from them.
\begin{lemma}
\label{lem:invariant} 
 The IMGs of equivalent networks are isomorphic: 
$$\tuple{f,W} \sim \tuple{f',W} \implies G_f/W \simeq G_{f'}/W.$$
\end{lemma}
Consequently the interaction graphs of two equivalent networks for the sequential mode are isomorphic because 
 the IMG is the unsigned version of the interaction graph. Moreover, we can reckon the sign of the interactions of the interaction graph directly from the interaction of the other by the function $\mu$ (Lemma~\ref{lem:equiv-async} ):
\begin{equation}
\mu(a_i \stackrel{x}{\longrightarrow} a_j
,(p,\pi)) =a_{\pi(i)} \stackrel{x(-1)^{p_{\pi(i)}+p_{\pi(j)}}}{\longrightarrow} a_{\pi(j)}, \quad p \in \Bset^n, \pi \in \gsym{n}.
\end{equation}
$\mu$ is extended to interaction graph by applying the function on each arc of the graph. For example, the functions defined in Figure~\ref{fig:iso} are equivalent (with $p=(1,1,1)$ and $\pi=e$) with the same interaction graph.

\begin{lemma}
\label{lem:equiv-async}
Let $\tuple{f,\{\{a_i\}\}_{a_i \in A}},\tuple{f',\{\{a_i\}\}_{a_i \in A}}$ be two networks equivalent for the sequential mode with respect to the signed permutation $(p,\pi) \in SRM_1^n$, we have: $G_{f'}=\mu(G_{f},(p,\pi)).$ 
\end{lemma}
Thereby, the isomorphic cycles (\ie with the same agents up to a permutation by $\pi$) of $G_f$ and $G_{f'}$ are of the same sign but the sign of their interactions may differ (Corollary~\ref{cor:equiv-async}).
\begin{corollary}
\label{cor:equiv-async}
The isomorphic cycles by $\pi$ in $G_f$ and $G_{f'}$ are of the same sign.
\end{corollary}

\subsection{Equivalence through mode variation}
Although the equivalence closely depends on the mode, it may be preserved from a mode to another by mode \emph{embedding} (Definition~\ref{def:piembed}).
\begin{definition}
\label{def:piembed}
Let $W$ and $W'$ be two modes partitioning $A$, let $\pi \in \gsym{k}$ be a permutation, \emph{$W$ is $\pi$-embedded} into $W'$ if and only if: 
\begin{enumerate}
\item Each modality of $W$ are included in a modality of $W'$:

$\forall w \in W, \exists w' \in W': w \subseteq w'$; \label{cond1}

\item The image by $\pi$ of the modalities of $W$ included in the same modality of $W'$ are also included to the same modality  and conversely. 

$\begin{array}[t]{l}
\forall w_i, w_j \in W, \forall w'\in W', \forall w'' \in W': \\  \qquad
 w_i \cup w_j \subseteq w' \iff w_{\pi(i)} \cup w_{\pi(j)} \subseteq w''. 
\end{array}$ \label{cond2}
\end{enumerate}
\end{definition} 
For example $\{\overbracket{\{a_6\}}^1,\overbracket{\{a_5\}}^2, \overbracket{\{a_3,a_4\}}^{3},\overbracket{\{a_1,a_2\}}^4 \}\}$ is $\pi-$embedded to:

\noindent
 $\{\{a_1,a_2,a_5\},\{a_3,a_4,a_6\}\}$ with $\pi =(1 \, 2) (3 \, 4)$ or $\pi=e$.

\begin{lemma}
\label{lem:fun-equivalent}
 Two networks equivalent for a mode with respect to an isomorphism $(\beta,\pi)$ are also equivalent for any mode $\pi$-embedding this mode.
\end{lemma}
Hence, an equivalence found with the sequential mode also holds for any mode since the sequential mode is embedded in all modes. An equivalence found for any mode also holds for the parallel mode since the parallel mode embeds all modes.

As a particular result found in \cite{Jarrah2010} and generalized in this article, a function in conjunctive normal form is equivalent to another function where \textsc{or} operators are merely replaced by the \textsc{and} operators and conversely (Figure~\ref{fig:iso}) for any mode. Indeed, 
for sequential mode, when $p=(1,\ldots,1)$ corresponding to a complete negation of agents and $\pi$ is the identity ($\pi=e$), the following property can be deduced from~(\ref{eq:equiv-fun}) that $f'_{a_i}(s)=\lnot{f_{a_i}( \lnot{s})}, a_i \in A$ as the inverse of the negation is the negation itself. From this equation, by application of the Morgan law and change of variables, from a function in disjunctive normal form $f$ we obtain an equivalent function $f'$ available for any mode where \textsc{or} are exchanged by \textsc{and} and conversely. 

 The equivalence properties of interaction graphs based on model isomorphism have also been studied by \cite{Noual2012}. The author proves that the models of two functions with the same non-signed interaction graph where each non-cyclic path has the same sign (\ie canonical form) are isomorphic at the asymptote for the generalized mode.

\section{Conclusion}
Isomorphism on the models of dynamics formalizes the analogy on  Boolean networks. 
The study characterizes the group of isomorphisms preserving the mode from which we deduce an equivalence on networks based on the isomorphism of their model. It reveals that the equivalence strongly depends on the mode: two equivalent networks may loose the equivalence when the mode varies. In this context, we show that the equivalence is maintained providing a mode is embedded in another. Besides, we show that the equivalence also induces an equivalence on a structure called the interaction modal graphs of the networks. This property is completed by an equivalence of the sign of the cycle for the sequential mode.

Several perspectives are considered. The characterization of the group could be extended to groups of isomorphisms preserving the spectrum. If these groups coincide for sequential and parallel mode because the mode can be trivially deduced from the spectrum without ambiguity, they differ for the other modes. This perspective enables the definition of a larger class of behaviorally similar Boolean networks while remaining close to the initial parametrization of model. Another perspective is related to the canonical form of networks resulting to  a transform of the network to a canonical representation dynamically equivalent to the original one. The canonical form is considered as the representative of a class of networks.
 The issue is to deduce properties of the dynamics of the network from the canonical form. 

\pagebreak
\appendix

\begin{proof}[Proposition~\ref{prop:subgroup}]
A Boolean permutation $\beta \in \gsym{\Bset^n}$ preserves the mode $W$ if and only if: 
$\forall \mathcal M: \mode M =W \iff \mode \mathcal M^\beta=W.$

Let $\mathcal M$ be a model with a mode $W$, we prove that:
\begin{itemize}
\item the group is closed by composition. Let $\beta_1,\beta_2 \in \gsym \Bset^n$ be two Boolean permutations preserving the mode $W$, we have: 

$\mode \mathcal M^{\beta_1\beta_2} = \mode (\mathcal M^{\beta_1})^{\beta_2} = W$ since $\mode \mathcal M^{\beta_1} = W$ and $\beta_2$ also preserves the mode $W$.
\item the identity (neutral element) $e$ preserves the mode since 

$\mode \mathcal M^e = \mode \mathcal M = W$.
\item if $\beta$ preserves the mode $W$ then the inverse $\beta^{-1}$ also preserves the mode. 
Assume that $\beta^{-1}$ does not preserve a mode $W$, we have:

$W=\mode \mathcal (M^\beta
)^{\beta^{-1}} =  \mode \mathcal M$ and $\mode \mathcal M \neq W$ which is false by hypothesis.
\end{itemize} 
\qed \end{proof}

\begin{proof}[Lemma~\ref{lem:sequential-group}] 
Let $\mathcal M$ be a model with a sequential mode  $\{\{a_i\}\}_{a_i \in A}$, by definition of the mode for all states $s \in \Bset^n$ and all modalities $a_i \in A$:  $s \xevo{a_i} \tilde f_{a_i}(s)$ implies that $\operatorname{hd}(s,\tilde f_{a_i}(s))=1$. Thus, the arcs of the model are also arcs of the $n$ dimensional hypercube.

\noindent
$(\implies) \quad$ As the automorphisms of the hypercube map any arc to other arc of the hypercube, the condition of the sequential mode are preserved because a model with a sequential mode is always a sub-graph of the hypercube. Thus, the sequential mode is preserved by an automorphism of the hypercube for all models. 

\noindent
\medskip
$(\impliedby) \quad$ 
By definition an automorphism of the hypercube the Boolean permutation $\beta$ complies to the following property: 
$$\forall b_1,b_2 \in \Bset^n: \operatorname{hd}(b_1,b_2)=1 \iff \operatorname{hd}(b_1^\beta,b_2^\beta)=1.$$ 
Thus, if $\beta$ is not an automorphism of the hypercube then there is an arc $(b_1,b_2)$ such that: $\operatorname{hd}(b_1^\beta, b_2^\beta) > 1$, meaning that the mode is not preserved for the models because the transition $b_1^\beta \evo  b_2^\beta$ does not correspond to the state change of one agent only.

We conclude that the automorphisms of the hypercube are the isomorphisms preserving the sequential mode for all models. The automorphisms of the $n-$dimensional hypercube~\cite{Harary2000, Chen1993} is also  isomorphic to the group of the signed permutation of rank $n$ or the Hyperoctahedral group of rank $n$ and denoted $BC_n$.
\qed \end{proof}

\begin{proof}[Proposition~\ref{prop:srm}]
We prove that $KM_W$ is isomorphic to the graph resulting of the cartesian product of $k$ complete graphs $K_{2^m}$, $ (K_{2^m})^{\Box k}$. 

Without loss of generality, we assume that: $V(K_{2^m})=\Bset^m$. For simplicity, the modalities of $W$ will be arbitrary indexed, \ie $W=(w_i)_{1 \leq i \leq k}$. 

\medskip
\noindent
Let $\beta_W: \Bset^n \to \Bset^n$ be a function mapping the vertices of $KW_W$ to the vertices of $(K_{2^m})^{\Box k}$, $\beta_W$ is defined as follows:

$\forall b,b' \in \Bset^n: b^{\beta_W}=b' \iff \forall w_i \in W: 
b[w_i]=b'[1+(i-1)m,\ldots,i.m].$

\medskip
\noindent
 In other words, the Boolean vector $b$ is re-ordered such that $b[w_i]$ is mapped to a sub-vector of $b'$ starting at position $1+(i-1)m$, 
\medskip
\noindent
$\beta_W$ is a bijection because $\beta$ re-orders the $k$ sub-vectors of length $m$ and no sub-vectors are overlapped due to the fact that $W$ is a partition.

\medskip
\noindent
Now, let us check that: $KM_W \stackrel{{\beta_W}}{\simeq} (K_{2^m})^{\Box k}$. 

\noindent
 By definition, for all $(b_1,b_2) \in E(KM_W)$ there exists $w_i \in W$ such that: 
$$b_1[w_i] \neq b_2[w_i] \wedge b_1[A \setminus \{w_i\}]= b_1[A \setminus \{w_i\}].$$ 

\noindent
We prove that  $(b_1^{\beta_W}, b_2^{\beta_W}) \in E( (K_{2^m})^{\Box k})$. 
 Let $b_1'=b_1^{\beta_W}$ and $b_2' = b_2^{\beta_W}$,  we deduce that there exists an edge 
$(b_1',b'_2) \in E( (K_{2^m})^{\Box k})$ because $b_1'[1+(i-1)m,\ldots,im] \neq b_2'[1+(i-1)m,\ldots,im]$ by hypothesis and the $i^\text{th}$ graph in the product is a complete graph. Hence, we conclude that: 

\noindent
$ \forall (b_1,b_2) \in E(KM_W): (b_1^{\beta_W}, b_2^{\beta_W}) \in E( (K_{2^m})^{\Box k});$

\medskip
\noindent
As $\beta_W$ is a bijection, we conclude that $KM_W \stackrel{\beta_W}{\simeq} (K_{2^m})^{\Box k}$.
\qed \end{proof}

\begin{proof}[Lemma~\ref{lem:srm}]
From Proposition~\ref{prop:srm}, we deduce that for any regular mode $W$ of length $m$, $\aut(KM_W) \simeq \aut( (K_{2^m})^{\Box k})$.
Thus, we focus on the characterization of $\aut( (K_{2^m})^{\Box k})$. A complete graph is prime with respect to the Cartesian product beacuse there is no induced square ($4-$cycle)~\cite{Imrich2008}. Hence, $K_{2^m}$ is also relatively prime with itself. Accordingly, the following property holds (\cite{Hammack2011}~Corollary~6.12, p.~70): 
 $$\aut( (K_{2^m})^{\Box k}) \simeq \aut(K_{2^m}) \wr S_k.$$ 
As $\aut(K_{2^m}) \simeq S_{2^m}$ and $ S_{2^m} \simeq \gsym {\Bset^m}$, we deduce that: 
$$\aut( (K_{2^m})^{\Box k}) \simeq S_{2^m} \wr S_k \simeq \gsym {\Bset^m} \wr S_k.$$
In conclusion, for all models $\mathcal M$ such that the spectrum is $\spectrum \mode \mathcal M = k \bullet m $, $k = \frac{n}{m}, n=|A|$, we have:
 $$\aut(KM_{(\mode \mathcal M)}) \simeq \aut( (K_{2^m})^{\Box k}) \simeq S_{2^m} \wr S_k \simeq \gsym{\Bset^m }\wr S_k.$$
 As $\aut(KM_{(\mode \mathcal M)})$ defines the set of isomorphisms preserving the mode of $\mathcal M$, we conclude that: $SRM^n_m \simeq \gsym \Bset^m \wr S_\frac{n}{m}, 1 \leq m \leq n.$
\qed \end{proof}

\begin{proof}[Theorem~\ref{the:sm}]

Let $W$ be a mode partitioning the agents such that $\spectrum W = \{k_i \bullet m_i\}_{i \in \llbracket r \rrbracket}$, We define $W_i, i \in \llbracket r \rrbracket$ as regular sub modes of $W$, such that $\spectrum W_i = k_i \bullet m_i$ and $W = \bigcup_{i=1}^r W_i$. 
The graph encompassing all the state graphs according to the mode $W$ is the cartesian product of graphs related to sub-mode $W_i$:
$$KM_W= KM_{W_1} \Box \ldots \Box KM_{W_i} \Box \ldots KM_{W_r}.$$

\medskip 
\noindent
Proposition~\ref{prop:srm} demonstrates that $KM_{W_i}$ is isomorphic to $K_{2^{m_i}}^{\Box k_i}$. Hence, we deduce that $KM_W$ is isomorphic to:
 $$K = K_{2^{m_1}}^{\Box k_1} \Box \ldots \Box K_{2^{m_r}}^{\Box k_r}.$$ 

\medskip
\noindent
Now, we focus on the characterization of the automorphisms of $K$.
Two products of complete graphs $K_{2^{m_i}}^{\Box k_i}$ and $K_{2^{m_j}}^{\Box k_j}$ such that $m_i \neq m_j$ are relatively prime. Indeed, a complete graph is prime and then cannot be generated from a product of complete graphs. Hence, a product of complete graphs is relatively prime to a product of another complete graph if the cardinality of the vertices differ. Then, $K$ is defined by its prime decomposition of complete graphs. As a complete graph $K_{2^{m_i}}$ is isomorphic to itself only according to the decomposition and prime, two cartesian products of complete graphs of different cardinality are not isomorphic and relatively prime. We deduce that: 
$$ \aut(K) \simeq \aut(K_{2^{m_1}}^{\Box k_1}) \times \ldots \times \aut(K_{2^{m_r}}^{\Box k_r}).$$
(See \cite{Hammack2011} Theorem~6.10 and Theorem~6.13, pp.~68--69)
 
\medskip
\noindent
By application of Lemma~\ref{lem:srm}, we conclude that:
$$\aut(KM_W) \simeq \aut(K) \simeq \times_{i=1}^{r} \gsym \Bset^{m_i} \wr S_{k_i}.$$
\qed \end{proof}

\begin{proof}[Lemma~\ref{lem:invariant}]
First, let us  define the interactions for modalities of $W$  extending Definition~\ref{def:interaction}:
$$ w_i \longrightarrow w_j \eqdef 
	\begin{array}[t]{l}
	\exists s_1,s_2 \in \Bset^n: s_1[w_i] \neq s_2[w_i] \wedge s_1[A \setminus w_i] = s_2 [A \setminus w_i] \wedge \\
 \quad f_{w_j}(s_1) \neq f_{w_j}(s_2).
	\end{array}
$$

For two equivalent networks $N=\tuple{f,W}$ and $N'=\tuple{f',W}$,  there exists an isomorphism $\varphi=(\beta,\pi) \in SP_\spectrum{W}$ such that for all transitions of the model of $N$, $s \stackrel{w_i}{\longrightarrow} s'$ there exists exactly one transition $s^\varphi \stackrel{w_{\pi(i)}}{\longrightarrow} s'^\varphi$ in the model of $N'$ ( Definition\ref{def:equivnet} and Property~\ref{eq:equiv-iso} ). We prove that $\pi$ is the isomorphisms between IMGs of $f$ and $f'$. More precisely,  For all interactions $w_i \longrightarrow w_j$ of the IMGs of $f$ we prove that there exists an unique interaction $w_{\pi(i)} \longrightarrow w_{\pi(j)}$ in the IMG of $f'$ and conversely.  

\medskip 
By definition of interactions (Definition~\ref{def:interaction}), for all interactions $w_i \longrightarrow w_j$ of the IMG of $f$, we deduce that there exists two states $s_1,s_2$ such that:
\begin{enumerate}
\item $s_1[w_i] \neq s_2[w_i]$ \label{inv:a}
\item $s_1[A\setminus w_i] = s_2[A\setminus w_i]$ \label{inv:b}
\item $f_{w_j}(s_1) \neq f_{w_j}(s_2)$ \label{inv:c}
\end{enumerate}

By definition of $\varphi \in SP_{\spectrum{W}}$, the isomorphism acts locally on each modality, $\beta=(\beta_1,\cdots,\beta_{|W|})$, such that $\beta_i, 1 \leq i \leq |W|$ is a permutation acting on $s[w_i]$. Then the result is permuted with respect to the permutation on modalities ($\pi$). Hence, for all modalities $w \in W$ 
we deduce that: 
 \begin{enumerate}
\item $s_1^\varphi [w_{\pi(i)}] \neq s_2^\varphi[w_{\pi(i)}]$ because $\beta_i$ is a permutation  acting locally on  the part of $w_i$ and (\ref{inv:a}).
\item $s_1^\varphi[A\setminus w_{\pi(i)}] = s_2^\varphi[A\setminus w_{\pi(i)}]$ because $(\beta_1,\cdots, \beta_{i-1},\beta_{i+1}, \cdots, \beta_{|W|})$ are permutations and (\ref{inv:b}). 
 \item $f'_{w_{\pi(j)}}(s_1) \neq f'_{w_{\pi(j)}}(s_2)$ because $\beta_j$ is a permutation and (\ref{inv:c}).
\end{enumerate}
Hence, as $\pi$ is a permutation  we deduce that there exists an unique interaction $w_{\pi(i)} \longrightarrow w_{\pi(j)}$ in the IMG of $f'$. 

\medskip
\noindent
 Thus, we conclude that the IMGs of the two functions are isomorphic and $\pi$ is the isomorphism.
\qed \end{proof}

\begin{proof}[Lemma~\ref{lem:equiv-async}]
The proof is based on the definition of $f'_{a_i}$ from $f_{a_i}$ for all $a_i \in A$ leading to a relation between the two interaction graphs $G_f, G_{f'}$. the mode $W=\{\{a_i\}\}_{a_i \in A}$ is sequential. Without loss of generality, we assume that $\pi=e$.

\medskip
\noindent
From Equation~\ref{eq:action}, we deduce that two equivalent networks, $N=\tuple{f,W},N'=\tuple{f',W}$ such that $\mathcal M \models N$ and $\mathcal M^p \models N'$ are related by the following equation: $\tilde f'_{a_i}(s^p)= \tilde f_{a_i}(s)^p$ , for all ${a_i} \in A$ and for all $s\in \Bset^n$. In the context of the sequential mode with an action defined by a Boolean vector as a representation of a signed permutation: $s^p = s \lxor p$,
the equation becomes: $$\tilde f'_{a_i} (s \lxor p) = \tilde f_{a_i} (s) \lxor p.$$
Setting $s'=s \lxor p$ and by a change of variable ($s'=s$), $f'$ is defined as:
$$ \tilde f'_{a_i} (s)=\tilde f_{a_i} (s \lxor p) \lxor p.$$
because $s \lxor p \lxor p = s$.

 Now, the issue is to characterize the formula of the local evolution function $f'_{a_i}$ by considering the permutation.
 $f_{a_i}$ is a formula in disjunctive normal form (DNF) for all $a_i \in A$. By definition of $p$, we have:
\begin{equation}
\label{eq:fprime}
 f'_{a_i}(s)= \left \{\begin{array}{l l} 
 \lnot{f_{a_i}(s \lxor p)} & \text{if $p_i=1$} \\
													f_{a_i}(s \lxor p) & \text{otherwise}
													\end{array} \right.
\end{equation}
																																										
The negation of a formula in disjunctive normal form leads to a formula in conjunctive normal form (CNF) by application of the Morgan law where the literals are the negation of the disjunctive normal form (DNF).
Besides, since the conversion between CNF and DNF of a formula involves the distributivity and the associativity and does not require the application of the negation, the both representations of a formula have the same set of literals. Therefore, 
let $\operatorname{Lit}(\varphi)$ be the set of literals of formula $\varphi$ in DNF, (\eg$ \operatorname{Lit}( (a \land b) \lor (\lnot c \land \lnot b))= \{ a, b, \lnot c, \lnot b \}$ ), we have the following property: 
\begin{equation*}
\label{eq:lit}
\operatorname{Lit}(\lnot \varphi)= \lnot{\operatorname{Lit}(\varphi)},
\end{equation*}
 where $\lnot L = \{\lnot l_1, \ldots, \lnot l_n \}$.
	
\noindent
As, $\lnot{f_{a_i}}$ has the same variables than $f_{a_i}$ for all $a_i \in A$, we deduce that the interaction graph is the same, meaning the graphs differ by the sign of their interactions only.

\medskip
\noindent
Now, we focus on the rules modifying the signs.
 Let $a_j \stackrel{+}{\longrightarrow} a_i$ be a positive interaction of $G_f$, we deduce the following rules from Equation~(\ref{eq:fprime}): 

\begin{center}
\begin{tabular}{ c c c p{.5\textwidth}}
 $p_j$ & $p_i$ & $ \_ \; \in \operatorname{Lit}(f'_{a_i})$ & Comments\\
\hline
$0$ & $0$ & $a_j$ & no change since $b \lxor 0 =b, \forall b \in \Bset$\\
$0$ & $1$ & $\lnot{a_j}$ & since $p_i=1$ implies the negation of $f_{a_i}$ (Equation~\ref{eq:fprime}) \\
$1$ & $0$ & $\lnot{a_j}$ & since $s(a_j) \lxor 1 = \lnot{s(a_j)}$ \\
$1$ & $1$ & $a_j$ & since $f'_{a_i}$ is the negation of $f_{a_i}$ ($p_i=1$), but applied on the negation of $s(a_j)$ ($p_j=1$). Hence 
The literal related to $a_j$ occurring in $f'_{a_i}$ is $\lnot{\lnot{a_j}}=a_j$. \\
\hline
\end{tabular}
\end{center}

\medskip
A similar analysis can be developed for a negative interaction. The literals belonging to $\operatorname{Lit}(f'_{a_i})$ are then the negation of the literal occurring in the table. For similar reasons, no change occurs for the non-monotone interactions (sign = $0$).
Finally, the expression $x (-1)^{p_i+p_j}$, where $x \in \{-1,0,1\}$ is the sign and $p_i,p_j \in \{0,1\}$, characterizes these rules concisely. 

In conclusion, 
$f'_{a_i}$ is completely defined from $p$ and $f_{a_i}$ for all $a_i \in A$, leading to $G_{f'} = \sigma(G_f,p)$ where $\sigma$ operates on the interactions of $G_f$ as follows: $\sigma(a_i \stackrel{x}{\longrightarrow} a_j,p) =a_i \stackrel{x(-1)^{p_i+p_j}}{\longrightarrow} a_j, p \in \Bset^n$.
\qed \end{proof}

\begin{proof}[Corollary~\ref{cor:equiv-async}]
Given a cycle occurring in the two interaction graphs, $G_f,G_{f'}$ \wrt equivalent network for the sequential mode with functions $f,f'$, we define $(x_1,\ldots,x_{l})$ as the sequence of the signs for the cycle in $G_f$ by considering that each interaction of the cycle is of the form: $a_i \stackrel{x_i}{\longrightarrow} a_{(i \bmod l)+1}$, such that the agents $a_i, 1 \leq i \leq l$ belong the cycle. For the sake of simplicity we consider that $\pi=e$.

\medskip
\noindent
From Lemma~\ref{lem:equiv-async}, we deduce that $( x_1(-1)^{p_1 + p_2}, \ldots, x_l(-1)^{p_{l}+p_1})$ is the sequence of the signs for $G_{f'}$ where $p$ is the Boolean vector representing the isomorphism between the models. The sign of the cycle in $G_f$ is $X= \prod_{i=1}^l x_i$ and the sign of the cycle in $G_{f'}$ is $X'=\prod_{i=1}^l x_i(-1)^{p_i+p_{(i+1) \bmod l}}$. The expression of the sign of the cycle in $G_{f'}$ can be simplified leading to the following expression:
 $$ X'=(-1)^{2 \left(\sum_{i=1}^l p_i\right)}\left(\prod_{i=1}^l x_i \right)= \left((-1)^{2 \left(\sum_{i=1}^l p_i\right)}\right)X. $$
As $(-1)^{2 y} = 1, \forall y \in \Nset$, we conclude that $X'=X$. The signs of the both cycles are identical \qed 
\end{proof}

\begin{proof}[Lemma~\ref{lem:fun-equivalent}]
We prove that if $W'$ $\pi-$embeds $W$ then for two equivalent function with respect to $\varphi S_{\spectrum W}$, $\varphi$ is also an isomorphism of $SP_{\spectrum W'}$. Hence, we prove that $\varphi=(\beta,\pi)$ preserves the mode $W'$.

\paragraph{$\beta$ preserves the mode $W'$}
As $\beta=(\beta_1,\ldots,\beta_l), l =|W|$ and for all $w_i \in W$, each $\beta_i$ acts locally on $b[w_i]$, we deduce that if $w_i \subseteq w'_j$ $\beta_i$ also acts locally on $b[w'_j]$. As each $w_i \in W$ is necessary included in a modality $w'_j \in W'$, we deduce that $\beta$ does not change the mode. 

\paragraph{$\pi$ preserves the mode $W'$}
$\pi$ is a permutation of $SP_\spectrum{W'}$ if and only if $\pi$ is a permutation of modalities of $W'$ and the permutation is applied on modalities with the same cardinality.

\medskip
According to Definition~\ref{def:piembed}.\ref{cond1} and as $W$ and $W'$ form partitions of $A$, all the modalities of $W'$ correspond to a partition of modality of $W$. Beside the modalities of the image by $\pi$, $w_{\pi(i)}$ also form a partition of another modality (Definition~\ref{def:piembed}.\ref{cond2}). Hence, $\pi$ permutes modalities of $W'$.

Now let us check that $\pi$ is a permutation on modalities of $W'$ with the same cardinality.
As $\pi$ preserves the mode $W$ by hypothesis, we have $|w_i|=|w_{\pi(i)}|, \forall i \in \llbracket |W| \rrbracket$. Hence, let $w'_j \in W'$ be a modality such that, 
$\bigcup_{i \in I_j} w_i = w_j', I_j \subseteq \llbracket |W| \rrbracket$, as $\{w_i\}_{i \in I_j}$ is a partition of $w'_j$,
and $\pi$ is a permutation complying to rules of a $SP$ group mapping modalities on modalities of the same cardinality, we deduce that a modality of $W'$ mapped by $\pi$ to another modality of $W'$ have the same cardinality: 
$|w'_j|=|\bigcup_{i \in I_j} w_i|=|\bigcup_{i \in I_j} w_{\pi(i)}|=|w'_k|$. 

\medskip
As $\beta$ and $\pi$ preserve the mode $W'$, we conclude that $\varphi \in SP¨_{\spectrum W'}$. However, its representation as a pair $(\beta',\pi')$ is different. 
\qed \end{proof}

\pagebreak
\bibliographystyle{plain}
\bibliography{equivalence}	
\end{document}